\documentclass{elsarticle}
\usepackage[utf8]{inputenc}

\usepackage{times}
\usepackage{varwidth}
\usepackage{url}
\usepackage{amsmath}
\usepackage{amssymb}
\usepackage{flushend}
\usepackage{color}
\usepackage{graphicx}
\graphicspath{ {figures/} } 
\usepackage{caption}
\usepackage{subcaption}
\usepackage{algpseudocode}
\usepackage{algorithm}
\usepackage{amsthm}
\usepackage{hyperref}
\usepackage{mathtools}
\usepackage[utf8]{inputenc}
\usepackage[normalem]{ulem}

\usepackage{thmtools}
\usepackage{thm-restate}

\declaretheorem[name=Definition]{definition}

\declaretheorem[name=Lemma]{lemma}
\theoremstyle{definition}

\newcommand{\proba}[2][]{\mathbb{P}_{#1}\left[#2\right]}
\newcommand{\expect}[2][]{\mathbb{E}_{#1}\left[#2\right]}
\newcommand{\probacond}[3][]{\mathbb{P}_{#1}\left[\left.#2~\right|\,#3\right]}
\newcommand{\expectcond}[3][]{\mathbb{E}_{#1}\left[\left.#2~\right|\,#3\right]}

\newcommand{\comb}[2]{C_{#1}^{#2}}
\newcommand{\defineq}{\stackrel{\mathclap{\tiny\mbox{\text{def}}}}{=}}

\newcommand{\anyobsgraph}[2]{\rho_{#2}^{(#1)}(G, n_p)}
\newcommand{\anyobslocal}[3]{\rho_{#2,*}^{(#1)}(G, n_p, #3)}
\newcommand{\edgeobsgraph}[1]{\anyobsgraph{#1}{e}}
\newcommand{\edgeobsedge}[2]{\anyobslocal{#1}{e}{#2}}
\newcommand{\nodeobsgraph}[1]{\anyobsgraph{#1}{n}}
\newcommand{\nodeobsnode}[2]{\anyobslocal{#1}{n}{#2}}

\newcommand{\edgedeg}[1]{edgedeg_{#1}}
\newcommand{\avgprobafit}{\hat{\mu}}

\journal{Cell Patterns 4.1 (2023): 100662}

\begin{document}

\begin{frontmatter}

\title{Detrimental Network Effects in Privacy: A Graph-theoretic Model for Node-based Intrusions}

\date{\today}

\author[turing,imperial]{Florimond Houssiau\footnotemark[1]}
\author[northeastern]{Piotr Sapieżyński\footnotemark[1]}
\author[telaviv]{Laura Radaelli\footnotemark[1]}
\author[telaviv]{Erez Shmueli}
\author[imperial]{Yves-Alexandre de Montjoye\footnotemark[2]}

\affiliation[turing]{organization={The Alan Turing Institute},
            addressline={2QR, John Dodson House, 96 Euston Rd},
            city={London},
            postcode={NW1 2DB},
            country={United Kingdom}}
\affiliation[imperial]{organization={Imperial College London},
            addressline={Exhibition Rd, South Kensington},
            city={London},
            postcode={SW7 2BX},
            country={United Kingdom}}
\affiliation[northeastern]{organization={Northeastern University},
            addressline={360 Huntington Ave},
            city={Boston},
            postcode={02115},
            state={MA},
            country={USA}}
\affiliation[telaviv]{organization={Department of Industrial Engineering, Tel Aviv University},
            addressline={P.O. Box 39040},
            city={Tel Aviv},
            postcode={69978},
            country={Israel}}
\begin{abstract}
Despite proportionality being one of the tenets of data protection laws, we currently lack a robust analytical framework to evaluate the reach of modern data collections and the network effects at play. We here propose a graph-theoretic model and notions of node- and edge-observability to quantify the reach of networked data collections. We first prove closed-form expressions for our metrics and quantify the impact of the graph's structure on observability. Second, using our model, we quantify how (1) from 270,000 compromised accounts, Cambridge Analytica collected 68.0M Facebook profiles; (2) from surveilling 0.01\% the nodes in a mobile phone network, a law-enforcement agency could observe 18.6\% of all communications; and (3) an app installed on 1\% of smartphones could monitor the location of half of the London population through close proximity tracing. Better quantifying the reach of data collection mechanisms is essential to evaluate their proportionality.
\end{abstract}

\end{frontmatter}

\footnotetext[1]{These authors contributed equally.}
\footnotetext[2]{Lead and corresponding author, \url{deMontjoye@imperial.ac.uk}}

\section{Introduction}

For reasons ranging from the democratization of communication technologies~\cite{castells2011rise} to urbanization and rural exodus, we are today more connected than ever. The fraction of the population living in cities is increasing and projected to reach 68\% by 2050~\cite{unitedurbanization2018}, while the average worldwide degree of separation dramatically shrank from 6 steps in 1969 to 3.5 steps today~\cite{travers1969experimental,ugander2011anatomy, backstrom2012four, edunov2016three}. 
Many positive network effects arise from such connectedness, including increased average GDP~\cite{bettencourt2013origins,bettencourt2007growth} and number of patents produced per capita~\cite{carlino2007urban}, and an accelerated diffusion of information~\cite{pickard2011time,khatib2011crystal}.
Network science emerged as a field to study human and societies through the lens of our connections~\cite{lazer2009life}. 

This connectedness however impacts our privacy as, in modern systems, (1) data often relates to more than one person (e.g. a text sent between two people) and (2) people often collect or have access to data about their friends and people around them (e.g. bluetooth close-proximity data).
In both cases, the network effects associated with modern connectedness often strongly amplify the scope and privacy impact of data collections.
Both the Snowden revelations~\cite{eddington2019snowden} and Cambridge Analytica~\cite{lapowsky2019cambridge} leveraged these network effects.
Yet, despite these being two of the major privacy scandals of the last decade, we currently still lack the analytical tools to correctly assess the reach and therefore proportionality of modern data collections.

Proportionality of data collections is one the tenets of data protection laws such as the European Union's General Data Protection Regulation (GDPR) and Law Enforcement Directive (LED) and is an integral part of Privacy Impact Assessments (PIAs). The European Data Protection Supervisor (EDPS) for instance states in its recent ``Guidelines on assessing the proportionality of measures that limit the fundamental rights to privacy and to the protection of personal data'' that to evaluate the proportionality of data collections one must assess both the ``number of people affected'' and whether it ``raises ‘collateral intrusions’, that is interference with the privacy of persons other than the subjects of the measure''~\cite{EDPS2019}.
The importance of proportionality was further emphasized in the context of data collections by law enforcement agencies in the Opinion 01/2014 of the Article 29 Working Party which emphasize the need for ``the scope of the proposed measure [to be] sufficiently limited'' in particular with respect to the ``number of people affected by the measure''.

We here propose a graph-theoretic model to evaluate the reach and therefore proportionality of modern data collections and attacks.
We define a general model to measure the privacy loss for individuals when nodes in the network (the ``primary'' nodes) have their data, and data from their neighborhood, collected.
Such data collections include installing a malicious app, having loose permissions on Facebook, and surrendering one's phone data to the authorities.
We then define two metrics for the vulnerability of networks to these attacks, node- and edge-\textit{observability}, reflecting what an attacker can learn.
We prove closed-form expressions for the observability of networks. Finally, we show that (1) our model allows to correctly estimate the number of Facebook accounts observed by Cambridge Analytica in 2014, using only the degree distribution of the graph; (2) under current legislation, surveillance of 0.01\% random phones in a mobile phone network would allow a law-enforcement agency to tap 18.6\% of all communications occurring on that network; and (3) a software installed on 1\% of the smartphones in London would enable its developer to monitor the hourly location of half of the population through close-proximity tracing.

\section{Results}
\begin{figure}[ht]
    \centering
    \includegraphics[width=1\linewidth]{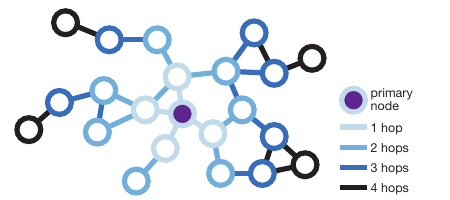}
    \caption{\textbf{Node- and edge-observability in a network with one primary node}. The direct neighbors (light blue) of the primary node, and the edges between the primary node and its neighbors are observable.
    The edges between the light blue nodes are however not observable.
    In the 2-hop case, nodes that are neighbors of nodes neighboring the primary node become observable. Similarly, all edges incident to a neighbor of the primary node become observable.
    }
    \label{fig:example_bob}
\end{figure}

We consider an undirected graph $G = (V,E)$, where $V$ is the set of nodes and $E$ is the set of undirected edges, i.e. unordered pairs $(u,v) \in E$ with $u,v \in V$. The edge-set of node $u$, the set of all its incident edges, is denoted by $E[u]$.
We define $V_p \subset V$, the set of \textit{primary nodes}, as the set that an attacker has gained control of through a node-based data collection or attack, e.g. a malicious app on Facebook or on a mobile phone. We denote this set's size by $n_p$.
This model applies to both attacks and legitimate data collections, and we use these words interchangeably.
An attacker can access all communication between any primary node $v \in V_p$ and its neighbors $\left\{u\,\left|\,(u,v) \in E[u]\right.\right\}$ in the network (phone calls, private messages, \dots), as well as any attributes of $v$'s neighbors that are available to $v$ (name, location, profile, \dots). This results in a set of observed edges $E_o$ and a set of observed nodes $V_o$. In some scenarios, the adversary may also observe nodes and edges located up to $k$ hops away from the primary node (see Fig.~\ref{fig:example_bob}).

\begin{definition}{\textbf{Observed Nodes}.}
The set of observed nodes after $k$ hops, $V_o^k$, for a set $V_p$ of primary nodes, contains all non-primary nodes that are at a shortest-path distance (noted $d$) at most $k$ from a primary node (where $d$ denotes the shortest path distance):
$$V_o^k \defineq \left\{ u \in V \setminus V_p: \exists\,v \in V_p, d(u,v) \leq k \right\}$$
\end{definition}

\begin{definition}{\textbf{Observed Edges}.}
The set of observed edges after $k$ hops $E_o^k$, for a set $V_p$ of primary nodes, contains all edges that are connected to at least one node within shortest-path distance $k-1$ of a primary node:

$$\begin{array}{ll}
    E_o^k \defineq & \left\{ (u,v) \in E : \exists w \in V_p,\right.  \\
    & \left.~~~~d(u,w)\leq k-1 \lor d(v,w) \leq k-1 \right\}
\end{array} $$

\end{definition}

\noindent For simplicity, we also define: $V_o \defineq V_o^1$
and $E_o \defineq E_o^1$. 

We then define two quantities to measure the reach of an attack, which we call edge-observability and node-observability. Edge-observability quantifies the amount of information transmitted between nodes accessible to an attacker; while node-observability quantifies the information about node attributes that this attacker has access to.
We also differentiate between two different risks in both attack models: the privacy loss caused by an attack on the whole network, $\anyobsgraph{k}{\circ}$), and by attacks on one particular node $u$, $\nodeobsnode{k}{u}$, or edge $e$, $\edgeobsedge{k}{e}$.

Finally, we focus here on non-targeted node-based intrusions as, in many real-world scenarios including the ones we consider here, such data collections or attacks are opportunistic rather than targeted: people install an app, click on a phishing link and enter their password, or pass by a malicious router in the street. In both attacks, the set $V_p$ is thus chosen uniformly at random: the probability for a node to be part of the data collection is uniform. Non-random attacks, when appropriate, can however be empirically studied using the same model. We discuss this in the Supplemental Information (section S1).

As all of our metrics depend on the number of primary nodes $n_p$, we also introduce the average node- and edge-observability (abbr. ANO, AEO), defined as $\frac{1}{N}\sum_{n_p=1}^N f(G, n_p)$ for a metric $f$. This number summarizes the observability curve and can be used as a measure of how vulnerable the network is to node-intrusion attacks in general.

\subsection*{Node-Observability}

The node-observability of a graph quantifies an attacker's ability to obtain knowledge about the attributes of a node through their relationship with primary nodes. 
Such attacks could, for instance, be conducted by malicious apps on social networks exploiting loose permissions to obtain the installer's friends information~\cite{venkatadri1investigating}.

\begin{definition}{\textbf{Probability of observation of a node.}}
The $k-$hop probability of observation of a node $u$, in a graph $G = (V,E)$, for $n_p$ primary nodes, $\nodeobsnode{k}{u}$, is defined as the probability that $u$ is observed, with the probability taken over all sets $V_p$ of size $n_p$ such that $u \not\in V_p$:
\begin{align*}
\nodeobsnode{k}{u} = \proba[\sim V_p]{u \in V_o^k~|~u \not\in V_p~\land~|V_p| = n_p}
\end{align*}
\end{definition}

\begin{definition}{\textbf{Node-Observability of a graph.}}
The $k-$hop node-observability of a graph $G = (V,E)$ for $n_p$ primary nodes, $\nodeobsgraph{k}$, is defined as the average fraction of nodes an attacker can observe, where the expectation is taken over all sets $V_p$ of size $n_p$:
\begin{align*}
    \nodeobsgraph{k} = \expect[\sim V_p]{\left.\frac{|V_p|+|V_o^k|}{|V|}~\right|~|V_p| = n_p}
\end{align*}
\end{definition}

These two quantities are linked one to another, as shown in theorem \ref{prop:prop_node_link}.

\begin{restatable}{theorem}{propnodelink}
\label{prop:prop_node_link}
Let $G=(V,E)$ be a graph with $n = |V|$ nodes, and $k$ a positive integer. The $k-$hop node-observability of $G$ for $n_p$ primary nodes is equal to:
$$\nodeobsgraph{k} = \frac{n_n}{n} + \frac{n-n_p}{n} \cdot \frac{1}{n} \sum_{u \in G} \nodeobsnode{k}{u}$$
\end{restatable}

Theorems \ref{prop:node-obs-graph} and \ref{prop:node-obs-node}, which we propose and prove, give provable closed-form expressions for the probability of observation of a node and the node-observability of a graph. These theorems rely on the $k-$hop degree of a node, i.e. the number of other nodes that are within at most $k$ hops from that node\footnote{In particular, note that the $1-$hop degree of a node is its degree.}.
In the 1-hop case, these expressions can be used to estimate node-observability and probability of observation, using only the full degree list of the graph. This quantity has been extensively studied and reported in the network science literature, see e.g. \cite{barabasi1999emergence, watts1998collective}.

\begin{restatable}{theorem}{propnodeobsgraph}
\label{prop:node-obs-graph}
Let $G = (V,E)$ be a graph with $n = |V|$ nodes, and $n_p \leq n$ a number of primary nodes. Let $u \in V$ be a node, the $k-$hop probability of observation of $u$ is given by:
$$\nodeobsnode{k}{u} = \left\{\begin{array}{ll}
    1 - \frac{\comb{n_p}{n-1-deg_k(u)}}{\comb{n_p}{n-1}} & \text{ if } deg_k(u) \leq n - 1 - n_p \\
    1 &  \text{ otherwise.}
\end{array}\right.$$

Where $deg_k(u)$ is the $k-$hop degree:
$$deg_k(u) = \left|\left\{v \in V \backslash \{u\}:~d(u,v) \leq k\right\}\right|$$
\end{restatable}

\begin{restatable}{theorem}{propnodeobsnode}
\label{prop:node-obs-node}
Let $G = (V,E)$ be a graph with $n = |V|$ nodes, and $n_p \leq n$ a number of primary nodes. Let $d \in \mathbb{N}^{n}$ be the distribution of $k-$hop degrees, defined as $d_i = \left|\left\{u \in V: deg_k(u) = i-1\right\}\right|$.
The $k-$hop node-observability of $G$ is given by:
$$\nodeobsgraph{k} = \frac{n_p}{n} + \frac{n - n_p}{n} \sum_{i=1}^n d_i f(n_p, i-1)$$
Where $f$ is defined as:
$$f(n_p, d) = \left\{\begin{array}{ll}
    1 - \frac{\comb{n_p}{n-1-d}}{\comb{n_p}{n-1}} & \text{ if } d \leq n - 1 - n_p \\
    1 &  \text{ otherwise.}
\end{array}\right.$$
\end{restatable}

\subsection*{Edge-Observability}

The edge-observability of a graph quantifies the attacker's ability to obtain edge-based information happening in the network when the data collector has access to $n_p$ random primary nodes.
Edge-observability can, for instance, be used to model the ability of a government agency to surveil communications occurring on a phone network.

\begin{definition}{\textbf{Probability of observation of an edge.}}
The $k-$hop probability of observation of an edge $e = (u,v)$, in a graph $G$, for $n_p$ primary nodes, $\edgeobsedge{k}{e}$, is defined as the probability that $e$ is observed, with the probability taken over all sets $V_p$ of size $n_p$: 
\begin{align*}
    \edgeobsedge{k}{e} = \probacond[\sim V_p]{e \in E_o^k}{|V_p| = n_p}
\end{align*}
\end{definition}

\begin{definition}{\textbf{Edge-Observability of a graph.}}
The $k-$hop edge-observability of a graph $G$ for $n_p$ primary nodes, $\edgeobsgraph{k}$, is defined as the expected fraction of the edge-set of a node $u$ that the attacker can observe, where the expectation is taken over all sets $V_p$ of size $n_p$: 
\begin{align*}
        \edgeobsgraph{k} = \expectcond[\sim V_p]{\frac{|E_o^k|}{|E|}}{|V_p|=n_p}
\end{align*}
\end{definition}

In theorems \ref{prop:edgeobsedge} and \ref{prop:edgeobsgraph}, we prove closed-form expressions to estimate the edge-observability of a graph and the probability of an observation of an edge in the $k$-hop case, using the notion of $k-$hop edge-degree, the number of nodes within distance $k-1$ of either extremity of the edge.
Importantly, we furthermore show that the average edge-observability can be computed from the number of nodes only (in the 1-hop case).

\begin{restatable}{theorem}{propedgeobsedge}
\label{prop:edgeobsedge}
Let $G = (V,E)$ be a graph with $N$ nodes, and $n_p \leq N$ a number of primary nodes. The $k-$hop edge-observability of $G$ is equal to the average $k-$hop probability of observation of edges $e \in E$.
$$\edgeobsgraph{k} = \frac{1}{|E|}\sum_{e \in E} \edgeobsedge{k}{e}$$
\end{restatable}

\begin{restatable}{theorem}{propedgeobsgraph}
\label{prop:edgeobsgraph}
Let $G = (V,E)$ be a graph with $n = |V|$ nodes, and $n_p \leq n$ a number of primary nodes. Let $e = (u,v) \in E$ be an edge. The $k-$hop probability of observation of $e$ is given by:
$$\edgeobsedge{k}{e} = \left\{\begin{array}{ll}
    1 - \frac{\comb{n_p}{n-\edgedeg{k}(e)}}{\comb{n_p}{n}} & \text{ if } \edgedeg{k}(e) \leq n - n_p \\
    1 & \text{ otherwise} 
\end{array}\right.$$

Where $\edgedeg{k}(e)$, for $e=(u,v)$, is the $k-$hop edge-degree of $e$:
$$\edgedeg{k}(e) = \left|\left\{w \in V:~d(u,w) \leq k-1 \lor d(v,w) \leq k-1\right\}\right|$$
\end{restatable}

\begin{restatable}{corollary}{corollaryedgeobsgraph}
\label{corollary:edgeobs-deg1}
Let $G = (V,E)$ be a graph with $n = |V|$ nodes, and $n_p \leq n$ a number of primary nodes. Let $e = (u,v) \in E$ be an edge. The $1-$hop probability of observation of $e$ is identical for all edges, and given by:
$$\edgeobsedge{1}{e} = 1 - \frac{n-n_p}{n} \cdot \frac{n - n_p - 1}{n - 1}$$
\end{restatable}

\subsection*{Impact of graph structure}
\label{sec:simulations}

In this section, we empirically investigate how the structure of the graph impacts its vulnerability to node-based data collections in the 2- and 3-hop case. We measure observability for four families of synthetic graphs: complete graphs, Erd\H{o}s-R\'{e}nyi~\cite{erdos1960evolution}, Barab\'{a}si-Albert~\cite{barabasi1999emergence} and Watts-Strogatz~\cite{watts1998collective} graphs. Each graph is generated with comparable number of nodes and edges (except for the complete graph, that always has $|E| = \frac{|V|(|V|-1)}{2}$), but exhibit different structure (see Section~\ref{sec:mm:empirical_obs}).

\begin{figure}[ht]
    \centering
    \includegraphics[width=1.0\textwidth]{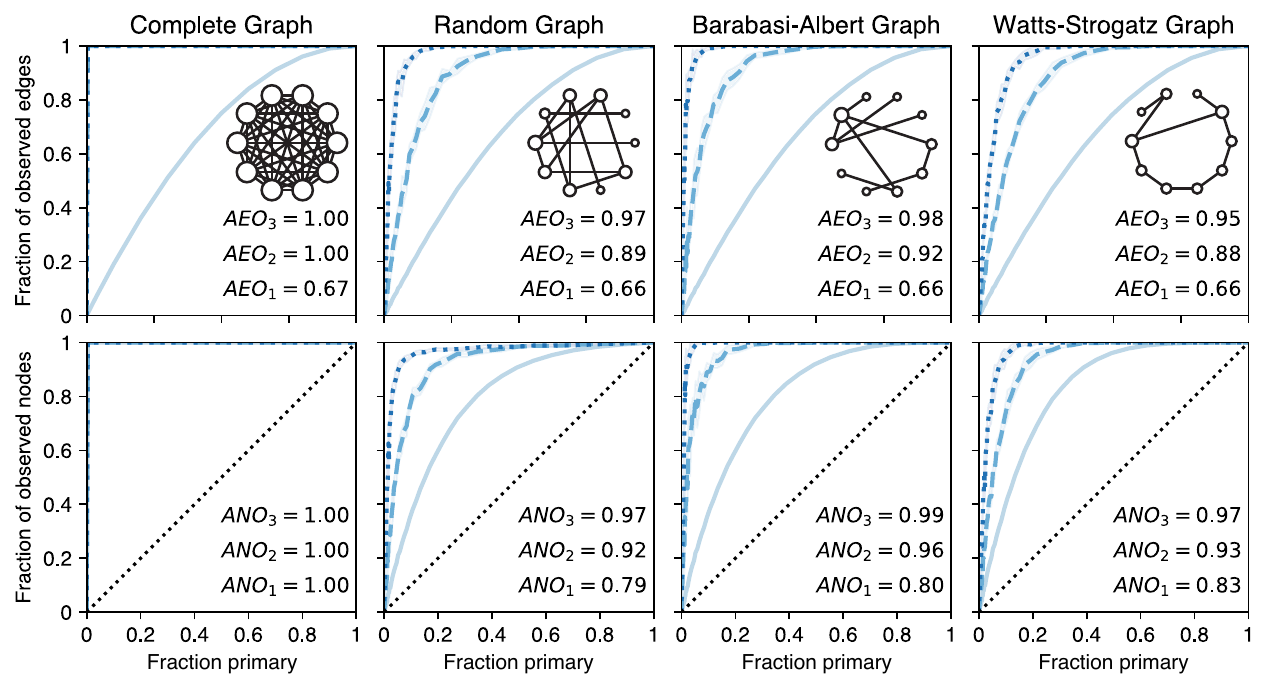}
    \caption{\textbf{$\mathbf{1-}$hop (light blue), $\mathbf{2}-$hop (dashed, blue) and $\mathbf{3-}$hop (dotted, dark blue) edge- and node-observability of synthetic graphs as a function of the fraction of primary nodes.}
    For each graph, we report the average node-observability (ANO) and average edge-observability (AEO).
    }
    \label{fig:observability_simulation_all}
\end{figure}

\begin{figure}[ht]
    \centering
    \includegraphics[width=1\linewidth]{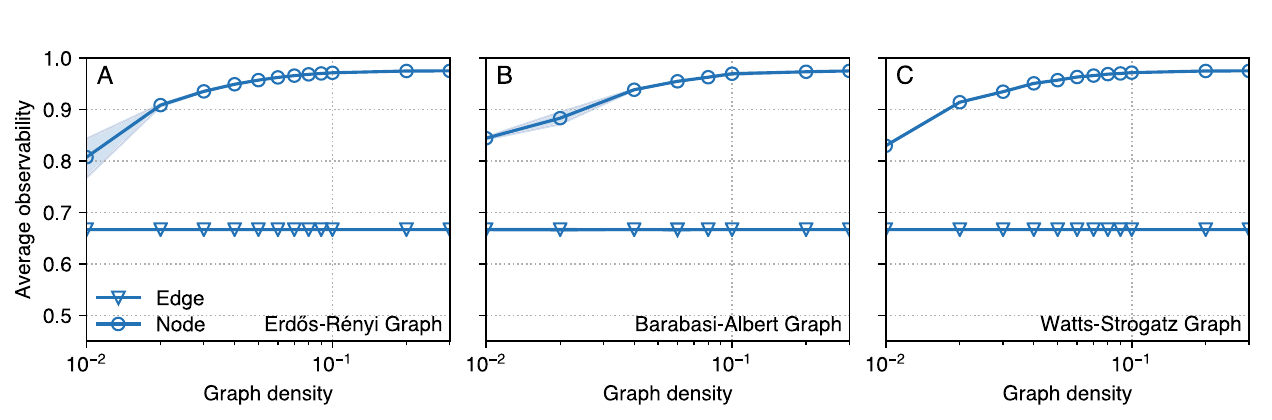}
    \caption{\textbf{Average $\mathbf{1-}$hop edge- and node-observability of (A) Erd\H{o}s-R\'{e}nyi, (B)  Barab\'{a}si-Albert, and (C) Watts-Strogatz graphs with increasing density.}
    The $1-$hop average edge-observability is constant (thm~\ref{prop:edgeobsgraph}), while the $1-hop$ average node-observability increases sharply with density.
    Barab\'{a}si-Albert graphs tend to be more observable than others, probably due to the presence of hubs.}
    \label{fig:density_all}
\end{figure}

First, we validate our theoretical results for 1-hop. We show in Fig.~\ref{fig:observability_simulation_all} that (1) the 1-hop edge-observability is identical for all types of graphs, as expected from theorems \ref{prop:edgeobsedge} and \ref{prop:edgeobsgraph}; (2) the node-observability and the average probability of observation of a node are near identical for all number of hops and small values of $n_p$, as expected from theorem \ref{prop:prop_node_link}.
Fig.~\ref{fig:density_all} shows that the 1-hop average edge-observability is constant at $2/3$ for increasing graph density, a direct consequence of theorem~\ref{corollary:edgeobs-deg1}. It further displays the sharp increase of the 1-hop average edge-observability with density.

Second, Fig.~\ref{fig:observability_simulation_all} shows the impact of the structure of a graph on its 2- and 3-hop node- and edge-observability.
More specifically, Barab\'{a}si-Albert graphs have larger average observability (for all metrics) than Erd\H{o}s-R\'{e}nyi graphs, which in turn have slightly higher average observability than Watts-Strogatz graphs.
This is likely due to Barab\'{a}si-Albert graphs containing, by design, very high-degree nodes called hubs. Their high degree makes hubs more likely to be observed which, in turn, allows an attacker to observe their neighbors (2-hop) and the neighbors of their neighbors (3-hop), thereby strongly increasing the observability of the graph.
Watts-Strogatz graphs, on the other hand, present a lattice-like structure with high clustering coefficient \cite{watts1998collective}.
Observing data from primary nodes is therefore likely to have a fairly local impact on observability, as opposed to spanning over the entire graph, hence decreasing the node- and edge-observability of the graph.

\subsection*{Social Networks}
\label{sec:social-networks}

In 2018, it was revealed that Cambridge Analytica, a UK company, had obtained data from about 70M US Facebook profiles through an innocuous-looking app. Once installed, the app would collect the data of the person who installed it as well as information about their Facebook friends (1-hop)~\cite{cadwalladr2018revealed,facebookprivacy2018}.
This led to investigations by several governments including the UK and the US, and resulted in Facebook receiving a record \$5 billion fine from the FTC~\cite{facebookfine2019}. This was, to the best of our knowledge, the first major instance of a node-observability attack based on node-intrusions on social networks.

\begin{figure}[ht]
    \centering
    \includegraphics[width=1\textwidth]{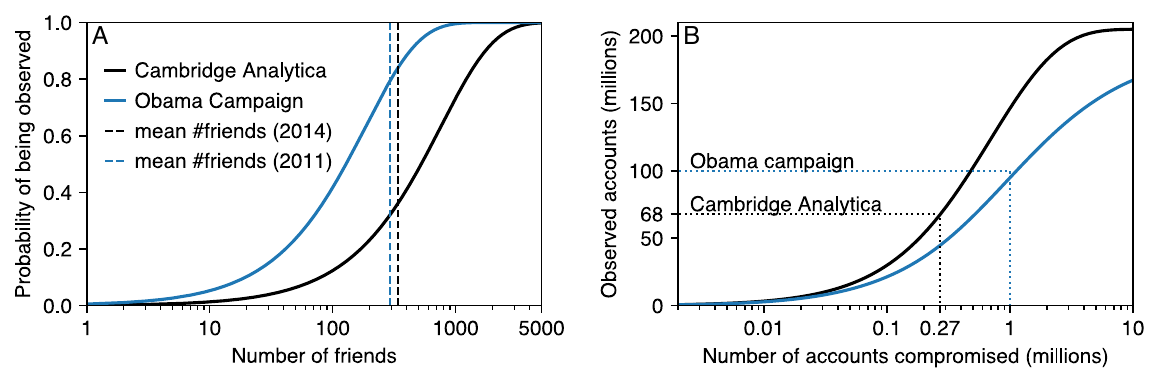}
    \caption{\textbf{Node-observability of the Facebook network in 2014}: \textbf{(A)} Probability of a node being observed by Cambridge Analytica or the Obama campaign as a function of the node's degree.
    The vertical dashed line is the average number of friends in 2014. \textbf{(B)}: node-observability (in absolute numbers) of the network as a function of the number of primary nodes (accounts installing the app), for 1- and 2-hops.}
    \label{fig:cambridge_analytica}
\end{figure}

Using our model, theorem~\ref{prop:node-obs-graph} and the shifted degree distribution of Facebook users in the United States~\cite{ugander2011anatomy}, we were able to quantify the 1-hop node-observability of the Facebook network. Fig.~\ref{fig:cambridge_analytica}B shows that, for the reported 270,000 app installs, the attacker should have been able to observe 68.0 million profiles ($\nodeobsgraph{1} = 0.318$). This contradicts initial reports that data about 50 million people had been breached, and is very close to later reports by Facebook that 70.6M US accounts had been breached~\cite{facebookprivacy2018}.
Similarly, our results show that, with 1M reported app installs, the Obama campaign might have had access to data of up to 94.8~million people~($\nodeobsgraph{1} = 0.510$).

Theorem~\ref{prop:node-obs-node} furthermore allows us to derive $\nodeobsnode{1}{u}$, the probability of a person to have had their data collected, as a function of their degree $deg(u)$. Fig.~\ref{fig:cambridge_analytica}A shows that an average user had a 36\% chance of having their profile collected by Cambridge Analytica and 79\% by the Obama Campaign. Users with as little as 100 Facebook friends would have had a 12.3\% (resp. 41.7\%) risk of having their profile collected by Cambridge Analytica (resp. the Obama campaign) while these numbers increase to 48.2\% (resp. 93.2\%) for users with 500 friends.

Finally, we evaluate the effectiveness of these attacks under a 2-hop policy, i.e. allowing apps to collect data on not only the friends, but also the friends of friends, of accounts installing them. 
Using a synthetic Facebook graph based on the configuration model \cite{barabasi2016network} and the known degree distribution, we show that a 2-hop policy would have allowed both Cambridge Analytica and the Obama campaign to observe virtually every single US profile.
We provide more details on our experimental methodology in the Supplemental Information (Section S2).

Note that although the error we observe between the node-observability of the model and the value published is relatively low ($3.7\%$), this is only one data point.
Our estimator has two main sources of error: (1) the estimation of the degree distribution from 2011 data, and (2) whether our assumption that primary nodes are sampled uniformly at random holds.
Further work is needed to quantify the impact the different sources of error have on the accuracy of the model. This is however challenging as an empirical analysis of error (2) would require data from many real-world attacks.

\subsection*{Mobile Phone Networks}
\label{sec:phone_network}

Surveillance of mobile phone networks was one of the first and most significant revelations by whistleblower Edward Snowden in 2013. The list of who we talked to on the phone---the edges of our social graph---is sensitive and considered private by 63\% of Americans~\cite{BCG2014}.
Recognising that this information might help fight crimes and terrorism, legislation allowing law enforcement to access a suspect's mobile phone records have been enacted in the past two decades around the world~\cite{mccarthy2002usa,franceinter2018}.
Several of these legislations, including in the US, also allow agencies to collect the mobile phone records (edges) of people up to two hops from the suspects (contacts of contacts), effectively recording communications occurring along 3-hop edges. 
For instance, in an attempt to curb illegal immigration, the US Border Patrol was allowed to use call logs to confirm the legal status of their potential target and to search for other potential illegal immigrants among the target's contacts~\cite{homeland_security}.
Similarly, the USA PATRIOT act allowed intelligence agencies to collect phone records of people up to three hops from suspects. This number was reduced to two hops in 2015, following the Snowden revelations~\cite{mayer2016evaluating}.

We here use a real-world mobile phone dataset of 1.4 million people over one month to study and quantify the potential of node-based data collections for surveillance. In this attack, we consider two nodes to be connected to one another if they interacted (call or text) at least once in the time period studied.

\begin{figure}[ht]
    \centering
 \includegraphics[width=1\linewidth]{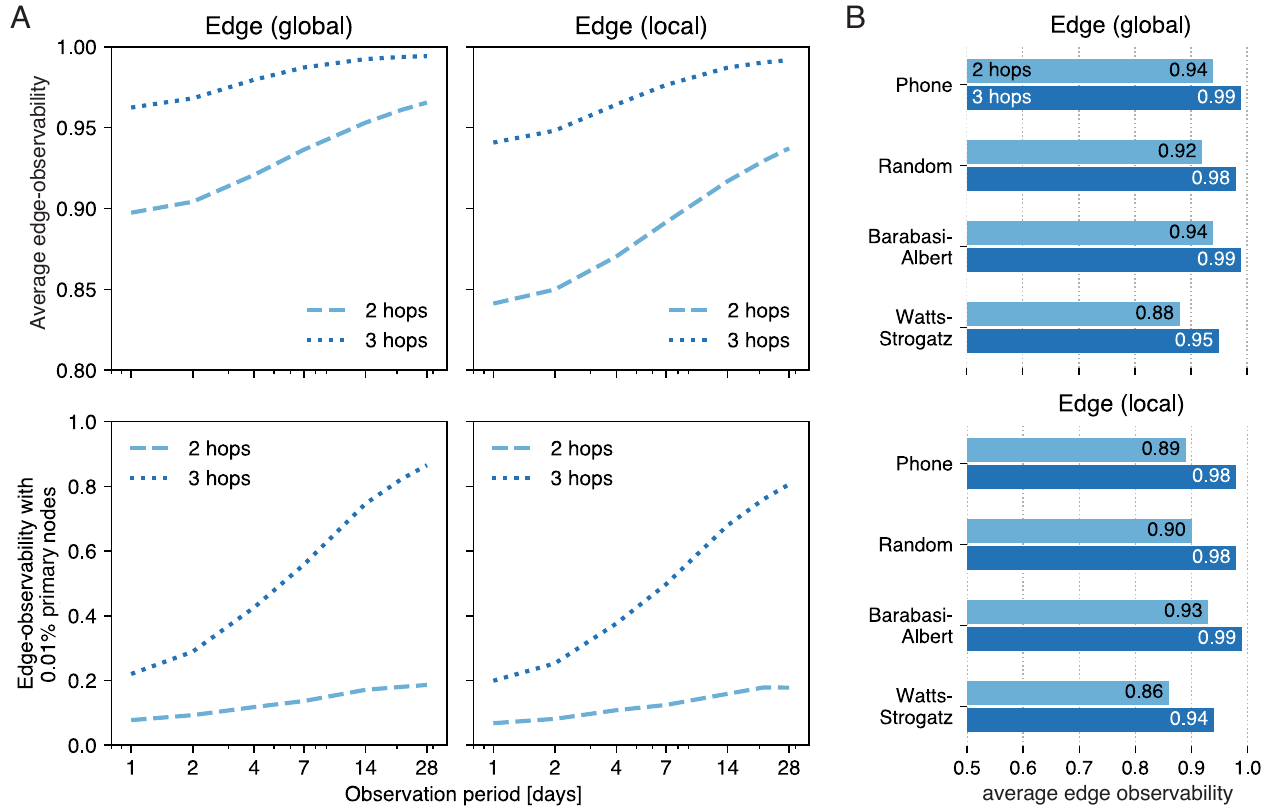}
    \caption{\textbf{$\mathbf{2-}$hop and $\mathbf{3-}$hop edge-observability of a mobile phone network.}
    \textbf{(A)} $2-$ and $3-$hop average edge-observability (top), and edge-observability when 0.01\% of the nodes are primary (bottom), of the real phone network as the observation window increases (in log-scale).
    \textbf{(B)} Comparison of the average edge-observability of the real phone network (observation window of 7 days) and synthetic networks with the same number of nodes and edges.}
    \label{fig:temporal_auc}
\end{figure}

Fig.~\ref{fig:temporal_auc}A shows that, under the previous 3-hop policy, collecting data from 0.01\% of the population would allow an attacker to surveil 86.6\% of all the communications happening in the network (edge-observability of the graph).
While the new 2-hop policy decreases those numbers, we found that it still allows an attacker to surveil 18.6\% of all the communications.
Finally, we show that the average edge-observability, when 1\% of the network is primary, (1) increases sharply with the length of the observation period and (2) already reaches 53.3\% when observing only one day of mobile phone metadata in the 3-hop case.

\begin{figure}[ht]
    \centering
    \includegraphics[width=1\linewidth]{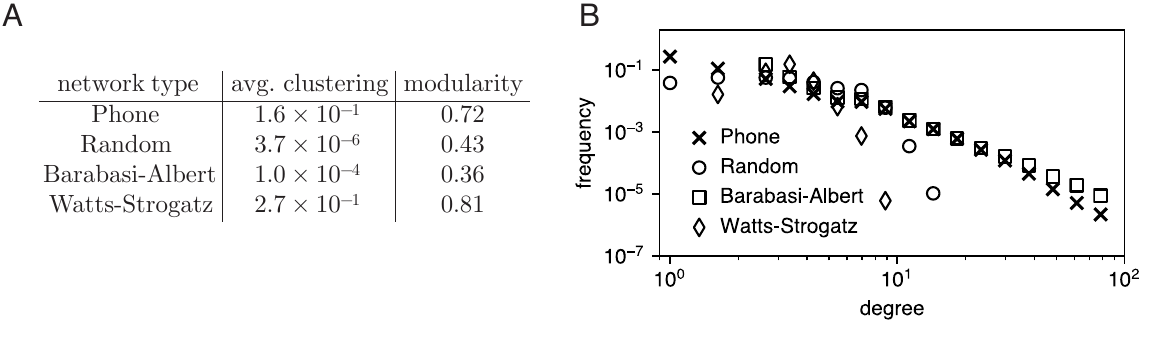}
    \caption{\textbf{Network properties of our phone network and three synthetic graphs}.
    \textbf{(A)} Modularity and average clustering coefficient for the mobile phone network and three synthetic graphs with identical number of nodes and density. Watts-Strogatz's resulting average clustering coefficient is the closest one to the one of our phone network. 
    \textbf{(B)} Degree distribution for the mobile phone dataset network and the synthetic graph models with equal number of nodes and density. The degree distribution of a  Barab\'{a}si-Albert graph appears to be the closest one to the empirical distribution.}
    \label{fig:network-properties}
\end{figure}

Fig.~\ref{fig:temporal_auc}B shows that the edge-observability of the phone network (7-day observation period) is comparable to that of a Barab\'{a}si-Albert graph with similar density and number of nodes.
As our previous results show, this is likely due to the presence of hubs in the mobile phone network.
Fig.~\ref{fig:network-properties} shows that both graphs have a similar degree distribution, including hubs, as is expected from previous research~\cite{aiello2000}.

\subsection*{Close proximity Networks}
\label{sec:colocation}

\begin{figure}[ht]
    \centering
    \includegraphics[width=1\linewidth]{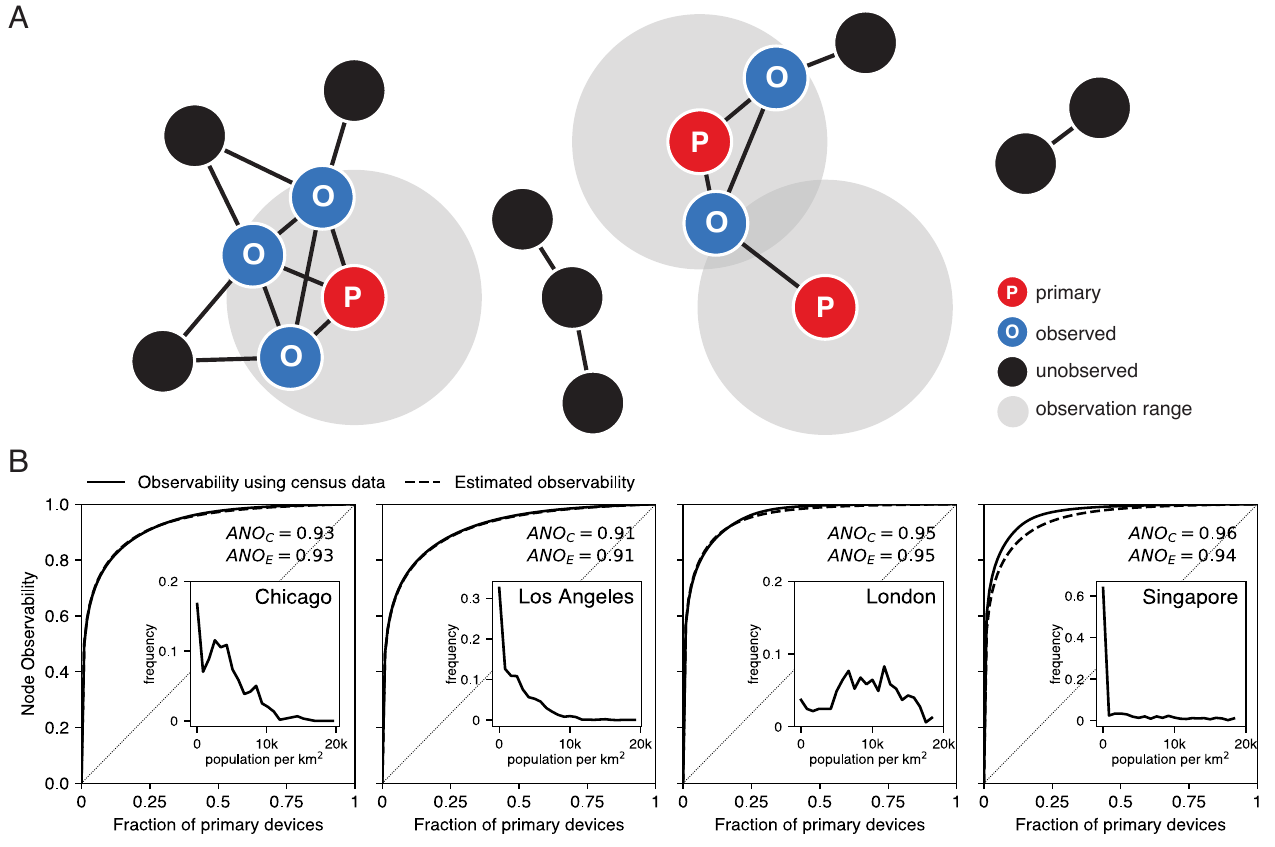}
    \caption{\textbf{Observability of a co-location network and estimation of the node-observability of cities.} \textbf{(A)} Compromised devices sense the presence of other devices within the observation range. This creates, for each time period, a co-location graph on which we compute local and global node observability. 
    \textbf{(B)} Node-observability in cities using exact census data ($ANO_C$) and estimation by an exponential distribution ($ANO_E$). Inset: distribution of the population per block of $1~km^2$ of each city.}
    \label{fig:dtu_attack}
\end{figure}

Mobility data is considered to be one of the most sensitive data currently being collected~\cite{BCG2014}.
We here use real-world Bluetooth close proximity data to investigate the feasibility and reach of a distributed node-observability attack through proximity sensing by compromised apps (e.g. a fake ``flashlight''~\cite{flashlight2017}) or contact tracing apps).
In this attack, an attacker monitors the location of nearby uncompromised devices using the GPS location and sensing capabilities of compromised (primary) phones.

\begin{figure}[ht]

    \centering
        \includegraphics[width=1\linewidth]{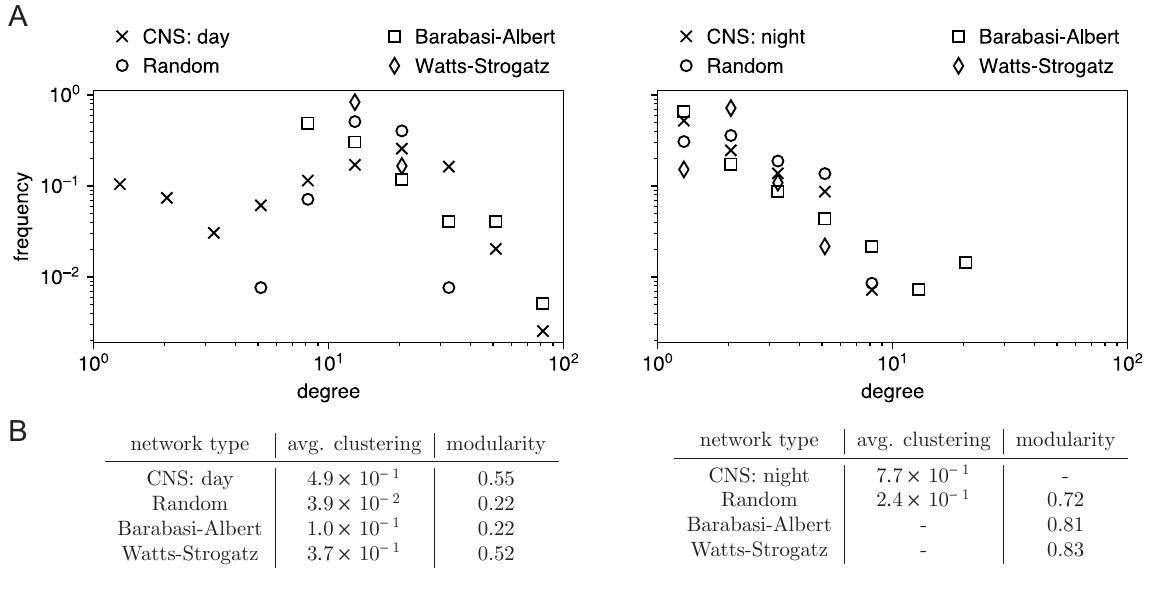}
        \caption{\textbf{Network properties of a daytime and a nighttime close proximity networks}.
        \textbf{(A)} Degree distribution for the Copenhagen Networks Study networks (a day snapshot and a night snapshot) and the synthetic graph models with equal number of nodes and density. The day snapshot has a bi-modal degree distribution that is not well reflected by any synthetic network. The night snapshot has a degree distribution similar to a Barabasi-Albert network.
        \textbf{(B)} Modularity and average clustering coefficient for Copenhagen Networks Study networks and three synthetic graphs with identical number of nodes and density. Watts-Strogatz's resulting average clustering coefficient is the closest one to day snapshot of CNS. The night snapshot is too sparse to meaningfully calculate clustering and modularity. Overall, the considered graph models are not a good fit for close proximity networks.
        }
        \label{fig:network-cns-properties}

\end{figure}

We empirically estimate the node-observability of this attack on a co-location dataset of 600 people collected as part of the Copenhagen Networks Study in 2014~\cite{stopczynski2014measuring}.
We extract hourly co-location graphs of phones who sensed each other within that hour, for different time periods and spatial regions (all of 1\,km$^2$, see Fig.~\ref{fig:dtu_attack}A).
From these hourly graphs, we compute the average probability of observation of a node as a function of the number of primary nodes per km$^2$, and fit an approximation curve $\avgprobafit(n_p) = \min\left(1, \max\left(0, A\log(n_p) + B\right)\right)$ ($R^2 = 0.876$).
From publicly available census data, we extract the population count in each 1~km$^2$ cell. We then combine these counts with $\avgprobafit(\circ)$ to estimate the node-observability of large cities (see Section~\ref{sec:mm:cities} and Supplemental Information, section S3 for more details).
In figure~\ref{fig:network-cns-properties}, we compare the network properties of two hourly graphs (one during daytime and the other during nighttime) with graph models. We show that the nighttime graph is well approximated by a Barab\'{a}si-Albert graph, while the daytime graph has a bimodal degree distribution that is not well approximated by the graph models we consider.

Fig.~\ref{fig:dtu_attack}B shows that using this attack in London, an attacker might be able to observe the location of as much as 57\% of individuals (resp. 86\%) with only 1\% (resp. 10\%) of the population installing the app, giving an average node-observability of 0.95. Other dense cities we considered -- Chicago, Los Angeles and Singapore -- display a similarly large ($> 0.876$) average node-observability.

Conversely, the attack is a lot less effective in sparsely populated areas where the average degree of the close proximity networks is lower. For instance, we estimate that in Utah ($15$ people per $km^2$ on average while London has $4500$) an app installed by 1\% (resp. 10\%) of the population would allow an attack to observe only 9\% (resp. 38\%) of individuals in the state, an average node-observability of $0.70$. 

Although the dataset we use is from 2014, we believe that our results still apply to modern datasets.
Indeed, we do not expect human mobility to have changed significantly (outside of the COVID-19 pandemic) between 2014 and now.
In the discussion, we further discuss how mobile phone applications could be leveraged to perform a similar attack.

\section{Discussion}
\label{section:Discussion}

Previous quantitative works have explored the impact of groups on individual privacy through the privacy loss incurred by homophily, the similarity between connected individuals in social networks~\cite{humbert2019survey} as the right of individuals to keep their affiliation with a group \cite{floridi2014open,Mislove2010,lampinen2015networked} -- a religion, a sexual orientation, a disease, etc. -- private. Researchers have then developed technical solutions to protect this affiliation (such as t-closeness~\cite{li2007t} or group-based obfuscation~\cite{narayanan2005obfuscated}).

Extensive prior work also studied the robustness of networks to node removal~\cite{callaway2000network}. Similarly to our work, nodes (and potentially their $l-$hop neighborhood~\cite{shang2011hop}) are compromised. However, the focus of such work is on estimating the impact of compromised node on the connectivity of the graph and its percolation clusters. Our work here focuses on the observability of graphs and its impact on privacy. 

In our mobile phone network experiment, we assumed that the data collector could gain access to a large number of primary nodes.
In practice, surveillance in phone networks is happening in numerous countries, sometimes at large scale. For instance, ANTENJ, a French national agency dedicated to ``numerical judiciary investigations'' has over 10000 phone numbers under surveillance at any point in time~\cite{franceinter2018}.
To the best of our knowledge, the only quantitative study of surveillance of phone networks has been conducted by Mayer et al.~\cite{mayer2016evaluating}. It concluded that hubs increase the number of nodes that can be reached through 3 hops. 
Their analysis is however entirely based on a disconnected dataset of less than 1000 users collected through an app. The small size of the dataset and the inherent sampling bias prevented them from 1) quantifying the potential of node-based intrusions for mass and targeted surveillance 2)  comparing their results to graph models, and 3) showing the importance of the graph's clustering coefficient in edge-observability.

In close proximity tracking attacks, phones could for instance be compromised by malware, code embedded by third-parties, or legitimate applications such as contact tracing apps. The Judy malware was for instance estimated to have infected 36 million devices worldwide~\cite{malware2017android}, while a fake flashlight app was discovered to collect data from tens of millions of users~\cite{ftc_flashlight}.
The UK start-up Tamoco, which reported having deals to embed their code in 1000 Android apps, giving them access to 100 million devices \cite{tamoco2018}, while the software of US company SignalFrame uses WiFi to scan for nearby devices and record their location~\cite{tau2020nextstep}.
Similarly, Pegasus, a spyware developed by NSO, has been employed to compromise the devices of targeted journalists, politicians, and activists in multiple countries~\cite{bergman2022pegasus}.
Finally, contact tracing applications such as those designed to measure the spread of COVID-19 are designed to track co-location of neighbouring phones~\cite{EuropeContactTracing2020} by having phones broadcast a unique identifier. If identifiers are consistent or linkable across time, they could be used for large-scale co-location tracing.

Close proximity data can be collected through Wi-Fi or Bluetooth.
Attacks based on Wi-Fi are either passive, where the app simply observes probe requests that mobile phones send to sense nearby Wi-Fi hotspots, or active, running a fake hotspot with a common SSID (e.g. \texttt{attwifi}, \texttt{xfinitywifi}) for nearby phones to connect to~\cite{jain2019unveil}. While more complex, the latter bypasses MAC address randomization used by some operating systems, including relatively recent versions of both Android and iOS~\cite{sapiezynski2015opportunities, sapiezynski2015tracking}. 
The two main platforms are thus moving towards re-randomization of the MAC address even when connecting to previously known networks~\cite{android2022}. These changes will however take years to reach all smartphone users.
Attacks based on Bluetooth, originally considered less likely, might be enabled by recent uses of Bluetooth for contact tracing apps against COVID-19~\cite{EuropeContactTracing2020}.
Notably, the use of frequent Bluetooth MAC address and broadcast payload randomization in these apps lowers the privacy threat. A motivated attacker may however still track either very partial location histories or recent histories of the users who report themselves COVID-positive~\cite{troncoso2020decentralized,dehaye2020proximity}.   

Close proximity sensing also presents an additional risk: primary users also reveal which other users they interact with, i.e. the edges of their close proximity network along with potentially other information such as the time of the interaction or distance between the two phones. This risk has been heavily discussed in the case of contact tracing apps relying on the so-called centralized model~\cite{dehaye2020proximity}. Corollary~\ref{corollary:edgeobs-deg1} shows that the resulting edge-observability (fraction of the edges observed) is independent from the graph's structure, being equal to $1 - \frac{n-n_c}{n}\cdot \frac{n-n_c-1}{n-1}\approx 1-(1-\frac{n_c}{n})^2$. This means that a contact tracing app which would report 15 days of a user's interactions to a server once that user is diagnosed positive would reveal, on average, 2.0\% of the 15 days social graph to the server if 1\% of the population is infected; and 18.9\% if 10\% of the population is infected. While network effects are still at play, they are a lot less strong than in the edge-observability case.

The attacks we model exploit the structure and connectedness of networks.
We therefore believe the main mitigation to be to limit access to the networks as described earlier in this section.
Another potential approach would be to modify the network in order to limit its observability.
We here study whether capping the maximum degree that nodes can have in a network reduces the risk.
We apply this cap to the social network setup (Section~\ref{sec:social-networks}), starting from 5000 (the maximum degree of that graph). We repeat our 1-hop analysis, measuring node-observability for $n_p = 270,000$, with a capped degree distribution $\hat{P}_\kappa$ obtained by counting all nodes with degree higher than $\kappa$ as nodes with degree $\kappa$ ($\hat{P}_\kappa(\kappa) = \sum_{i=\kappa}^{5000} \hat{P}_\kappa(i)$ and $\hat{P}_\kappa(i) = 0,\forall i > \kappa$).
We show in Table~\ref{tab:observability_ca_capped} that unless the cap is large enough to affect a significant fraction of all users ($> 20\%$), the node-observability of the network remains largely unchanged.
Protecting users from attacks by capping the maximum degree of nodes thus comes at a steep cost in utility, which might not be acceptable in practice.

\begin{table}[ht]
    \centering
    \begin{tabular}{ccc}
    \textbf{Degree cap} & \textbf{Fraction of nodes affected} & \textbf{Observability} \\ \hline
    5000 & 0.00\% & 0.329 \\
    4000 & 0.03\% & 0.329 \\
    2000 & 0.29\% & 0.329 \\
    1000 & 2.94\% & 0.326 \\
    500 & 17.21\% & 0.305 \\
    250 & 49.23\% & 0.243 \\
    \end{tabular}
    \caption{\textbf{Node-observability $\nodeobsgraph{1}$ for the 2014 Facebook graph and $n_p = 270,000$ accounts observed when the maximum degree is capped.} The observability only decreases slightly with the cap, except when a large fraction of nodes are affected.}
    \label{tab:observability_ca_capped}
\end{table}

This work is, to the best of our knowledge, the first to formalise and develop graph-theoretic models to understand the reach and therefore proportionality of modern data collection. Our results shed light on how network effects can have a strong detrimental impact on our privacy.
The reliance on network effects to collect data by Cambridge Analytica is one of the first large-scale examples of a node-based data collection and is unlikely to be the last.
We believe models like the one we present here to be essential to reason about data collection in networked environments.
Moving forward, we hope this work can help evaluate the scope of data collections mechanisms and technologies and ensure their proportionality.

\section{Experimental Procedures}
\label{section:MaterialsMethods}

\subsection{Resource availability}

\subsubsection{Lead contact}
Further information and requests for resources should be directed to and will be fulfilled by the lead contact, Yves-Alexandre de Montjoye (\url{deMontjoye@imperial.ac.uk}).

\subsubsection{Materials availability}
There are no physical materials associated with this study.

\subsubsection{Data and code availability}
Code to compute both edge and node observability and data will be made available at \url{https://github.com/computationalprivacy/network-privacy}.

The mobile phone data cannot be made available for privacy and confidentiality reasons. 
Instead, we will make available synthetic data for the mobile phone graph using the Barab\'{a}si-Albert graph model.
The social network data is available from Ugander et al.~\cite{ugander2011anatomy}. The Copenhagen dataset is available at Sapiezynski et al.~\cite{sapiezynski2019interaction}. Finally, we will make the population and area data about cities available.

\subsection{Proofs}
\label{sec:mm:proofs}

In order to prove the theorems in the paper, we introduce a useful lemma.

\begin{lemma} \label{lemma:subsetenum}
Let $G = (V,E)$ be a graph with $n = |V|$ nodes, $n_p \leq N$ a number of primary nodes, and $S \subset V$. The following holds when $V_p$ is drawn uniformly at random:
$$\probacond[\sim V_p \subset V]{\exists\, u \in V_p:~u \in S}{|V_p| = n_p} = \left\{\begin{array}{ll}
    1 - \frac{\comb{n_p}{n - |S|}}{\comb{n_p}{n}} & \text{ if } |S| \leq n - n_p  \\
    1 & \text{ otherwise}
\end{array}\right.$$
\end{lemma}
\begin{proof}
First, we rewrite the probability by considering its negation:
$$\begin{array}{ll}
    \probacond[\sim V_p \subset V]{\exists\, u \in V_p:~u \in S}{|V_p| = n_p} &= 1 - \probacond[\sim V_p \subset V]{\forall\, u\in V_p:~u \not\in S}{|V_p = n_p} \vspace{4pt} \\
     &= 1 - \probacond[\sim V_p \subset V]{V_p \cup S = \emptyset}{|V_p|=n_p} \vspace{5pt} \\
     &= 1 - \probacond[\sim V_p \cup V]{V_p \subset V \backslash S}{|V_p| = n_p}
\end{array}$$

Writing this probability in expectation of an indicator form yields:
$$\probacond[\sim V_p \subset V]{V_p \subset V \backslash S}{|V_p|=n_p} = \sum_{v_p \subset V, |v_p|=n_p} \proba{V_p=v_p} \cdot I\{v_p \subset V \backslash S\}$$

Since $V_p$ is drawn uniformly at random, and a set of size $n$ has $\comb{n_p}{n}$ subsets of size $n_p$, hence: $$\proba{V_p=v_p} = \frac{1}{\left|\left\{v_p \subset V: |v_p|=n_p\right\}\right|}$$

Furthermore, we have that, by definition:
$$\sum_{v_p \subset V, |v_p|=n_p} I\{v_p \subset V \backslash S\} = \left|\left\{v_p \subset V \backslash S: |v_p|=n_p\right\}\right|$$

The number of subsets of size $n_p$ of a set of size $m$ is $\comb{n_p}{m}$. Also, since that $S \subset V$, $|V \backslash S| = |V| - |S|$, and hence we obtain:

$$\probacond[\sim V_p \subset V]{V_p \subset V \backslash S}{|V_p|=n_p} = \frac{\left|\left\{v_p \subset V \backslash S: |v_p|=n_p\right\}\right|}{\left|\left\{v_p \subset V: |v_p|=n_p\right\}\right|} = \frac{\comb{n_p}{n - |S|}}{\comb{n_p}{n}}$$

\end{proof}

\propnodelink*

\begin{proof}
By definition and linearity of the expectation, we have (for the sake of clarity, we omit the conditional on $ |V_p| = n_p$ in expectations):
$$\nodeobsgraph{k} = \expect[\sim V_p]{\frac{|V_p|+|V_o^k|}{|V|}} = \frac{n_p}{n} + \frac{1}{n} \cdot \expect[\sim V_p]{|V_o^k|}$$

To compute the last term, we write $|V_o^k|$ as the sum of the indicator of whether a node is in $V_o$:
$$\expect[\sim V_p]{|V_o^k|} = \expect[\sim V_p]
{\sum_{u \in G} I\left\{u \in V_o^k\right\}} = \sum_{u \in G} \proba[\sim V_p]{u \in V_o^k}$$

To relate $\proba[\sim V_p]{u \in V_o^k}$ to $\nodeobsnode{k}{u}$, we develop it conditionally to the event $u \in V_p$, as, for any node $u \in G$:
$$\begin{array}{cl} \vspace{4pt}
    \proba[\sim V_p]{u \in V_o^k} = &  \probacond[\sim V_p]{u \in V_o^k}{u \in V_p}\proba[\sim V_p]{u \in V_p}+ \\
     &\probacond[\sim V_p]{u \in V_o^k}{u \not\in V_p}\proba[\sim V_p]{u \not\in V_p}
\end{array}$$

By definition, $V_p \cap V_o^k = \emptyset$, and thus the first term is zero. Since $V_p$ is a set of $n_p$ nodes selected uniformly at random from $n$ nodes, $\proba[\sim V_p]{u \not\in V_p} = \frac{n-n_p}{n}$. Hence:
$$\proba[\sim V_p]{u \in V_o^k} = \frac{n-n_p}{n} \cdot \nodeobsnode{k}{u}$$
\end{proof}

\propnodeobsgraph*

\begin{proof}

First, from the definition of $V_o$, we obtain:
$$\begin{array}{cl} \vspace{4pt}
 \nodeobsnode{k}{u} &=\probacond[\sim V_p\subset V]{u \in V_o^k}{u \not\in V_p,\,|V_p|=n_p}\\ \vspace{4pt}
 &=\probacond[\sim V_p\subset V]{\exists\,v \in V_p: d(u,v) \leq k}{u \not \in V_p,\,|V_p|=n_p}\\
 &=\probacond[\sim V_p\subset V]{\exists\,v \in V_p: d(u,v) \leq k}{V_p \subset V \backslash \{u\},\,|V_p|=n_p} \vspace{4pt}\\
 &=\probacond[\sim V_p\subset V]{\exists\,v \in V_p: v \in neigh_k(u)}{V_p \subset V \backslash \{u\},\,|V_p|=n_p}
\end{array}$$

\noindent Where $neigh_k(u) = \left\{v \in V \backslash \{u\}: d(u,v) \leq k\right\}$. Note that, $|neigh_k(u)| = deg_k(u)$.

\noindent Define $G' = (V',E') = (V \backslash \{u\}, E \backslash E[u])$, the graph without $u$. We can then write:
$$\begin{array}{cl}
 \nodeobsnode{k}{u} &=\probacond[\sim V_p \subset V']{\exists\,v\in V_p: v \in neigh_k(u)}{|V_p|=n_p}\\
\end{array}$$

\noindent Since $neigh_k(u) \subset V'$, we can apply lemma \ref{lemma:subsetenum} on $G'$ with $S = neigh_k(u)$ (observe that $|V'| = n-1$), which concludes the proof.
\end{proof}

\propnodeobsnode*

\begin{proof}
This follows immediately from theorems \ref{prop:prop_node_link} and \ref{prop:node-obs-graph}, using the fact that the probability of observation of a node depends only on its degree.
\end{proof}

\propedgeobsedge*

\begin{proof}
By definition, and rewriting $|E_o^k$ as a sum of indicator functions for elements of $E$:
$$\begin{array}{cl} \vspace{4pt}
    \edgeobsgraph{k} &= \expectcond[\sim V_p]{\frac{|E_o^k|}{|E|}}{|V_p|=n_p}  \\ \vspace{4pt}
     &= \frac{1}{|E|}\expectcond[\sim V_p]{\sum_{e \in E} I\left\{e \in E_o^k\right\}}{|V_p|=n_p} \\ \vspace{4pt}
     &= \frac{1}{|E|} \sum_{e \in E} \probacond[\sim V_p]{e \in E_o^k}{|V_p|=n_p}\\
     &=\frac{1}{|E|}\sum_{e \in E}\edgeobsedge{k}{e}
\end{array}$$
\end{proof}

\propedgeobsgraph*

\begin{proof}
We develop the probabilities, using the definition of $E_o$ and the definition of conditional probabilities:
$$\begin{array}{cl} \vspace{4pt}
    \edgeobsedge{k}{e} &= \probacond[\sim V_p]{e \in E_o}{|V_p| = n_p}  \\ \vspace{4pt}
    &= \probacond[\sim V_p \subset V]{\exists w \in V_p: d(u,w) \leq k-1 \lor d(v,w) \leq k-1}{|V_p|=n_p} \vspace{4pt}\\
    &= \probacond[\sim V_p \subset V]{\exists w \in V_p: w \in eneigh_k(e)}{|V_p| = n_p}
\end{array}$$

\noindent Where $eneigh_k(e) = \left\{w \in V: d(u,w) \leq k-1 \lor d(v,w) \leq k-1\right\}$. Note that $|eneigh_k(e)| = \edgedeg{k}(e)$, by definition. We then apply lemma \ref{lemma:subsetenum} for $G$ and $S = eneigh_k(e)$, which concludes the proof.
\end{proof}

\corollaryedgeobsgraph*
\begin{proof}
Observe that for $k=1$ and for any $e = (u,v) \in E$:
$$\edgedeg{1}(e) = \left|\left\{w \in d(u,w) \leq 0 \lor d(v,w) \leq 0\right\}\right| = \left|\{u,v\}\right| = 2$$
\noindent The result then follows from theorem \ref{prop:edgeobsedge}.
\end{proof}

\subsection{Datasets}
\label{sec:mm:datasets}

\textbf{Phone dataset:~~}
 Our phone dataset comprises four weeks of domestic intra-company communications phone logs (calls and texts) of 1.4 million customers of a mobile phone provider.
 
 \noindent
\textbf{Co-location dataset:~~}
  The dataset contains mobility information of about 600 students at a European university collected over a month (retrieved via GPS, Wi-Fi, or a combination of the two) along with Bluetooth sensing data (every 5 min), as part of the Copenhagen Networks Study~\cite{stopczynski2014measuring}.

\subsection{Empirical study of observability}
\label{sec:mm:empirical_obs}

We study graphs generated from four models: a complete graph and three random graphs with an average density of $0.015$; a Erd\H{o}s-R\'{e}nyi graph  (with $p=0.015$), Barab\'{a}si-Albert graph (with $m=2$), and a Watts-Strogatz graph (with $k = 5$ and $p = 0.2$), each with 250 nodes \cite{erdos1960evolution,barabasi1999emergence,watts1998collective}. We estimate the node- and edge-observability for each graph type, graph size, and number of primary nodes $n_p$ by selecting 500 random sets $V_C$ of $n_p$ nodes, measuring $\left|E_o^k\right|$ and $\left|V_o^k\right|$. The curves we report are the average over these sets.
To model the influence of graph density on observability (Fig.~\ref{fig:density_all}), we repeat the same procedure for densities ranging from $0.01$ to $0.3$, by changing the parameters $p, m$ and $k$ of the graph models. 

\subsection{Observability of cities}
\label{sec:mm:cities}

In order to estimate the average probability of observation as a function of the number of primary nodes for an area of 1~km$^2$, we compute hourly graphs from the colocation dataset (Fig.~\ref{fig:dtu_attack}A), and fit an approximation of the average probability of observation $\frac{1}{|V|}\sum_{u \in V}\nodeobsnode{k}{u}$ for $n_C$ primary nodes of the form $\avgprobafit(n_p) = 0.13 \log(n_p) - 0.05$~($R^2 = 0.876$). We use census data to compute population density in blocks of 1~km$^2$.

Finally, we use algorithm~\ref{algo:city} to estimate node-observability of a city when a fraction $x$ of the population is part of the primary data collection, with $n_B$ the number of blocks in a city, and $B$ a list of $n_B$ elements, the population size in each block (given by census data).
Algorithm~\ref{algo:city} estimates the average probability of observation in each block using the approximation $\avgprobafit(\circ)$, then scales it to obtain the node-observability (Thm.~\ref{prop:prop_node_link}).

\begin{algorithm}[ht]
\caption{Node-observability of a city}
\begin{algorithmic}[1]
\Procedure{NodeObsCity}{$n_B, B, x$}
\State $population \gets \sum_i B_i$
\State $observed \gets 0$
\For {$i = 1, \dots, n_B$}
    \State $m_i \gets x \cdot B_i$
    \State $observed~+=~\left(x + (1-x)\cdot \avgprobafit(m_i)\right) \cdot B_i$
\EndFor
\State \textbf{return} $observed/population$
\EndProcedure
\end{algorithmic}
\label{algo:city}
\end{algorithm}

Sometimes, detailed census data might not be available. Figure~\ref{fig:dtu_attack}C also shows that the node-observability in the cities we considered can be reasonably well approximated by using an exponential distribution for the distribution of the population per km$^2$ in the city (as an approximation of the actual distribution). The parameter $\lambda$ of the exponential is the density of the entire city $\lambda = \frac{\text{population}}{\text{area}}$. This allows for the global node-observability of a city --the fraction of the population observed by an attacker-- to be readily estimated knowing only the population and area of the city. 

This extrapolation from the CNS data to cities relies on two assumptions.
First, it assumes that the probability for an individual to be observed depends only on the absolute number of primary people it might encounter ($n_p$) in the cell, and not its total population.
Second, it assumes that the mobility of people in the CNS dataset resembles that of the population of cities. This doesn't account for differences in lifestyle, geography or population type.
However, we believe this to be the first attempt at estimating the risk of mass surveillance in large cities based on actual co-location data, and look forward to future work exploring this question.

\section{Acknowledgments}

We would like to thank Sune Lehmann for his invaluable insights. We would like to acknowledge Alan Mislove, Christo Wilson, Arnaud Tournier, Shubham Jain, Andrea Gadotti, and Ali Farzanehfar for their feedback.

\section{Author Contributions}

Conceptualization, all; Methodology, all; Software, LR and PS; Formal Analysis, FH; Investigation, LR, PS, FH and YAdM; Resources and PS, YAdM; Data Curation, PS and LR; Writing -- Original Draft, LR, PS, FH, and YAdM; Writing -- Review \& Editing, FH and YAdM; Visualization, PS and FH; Supervision, YAdM and ES; Project Administration, YAdM and FH.

\section{Declaration of Interests}

The authors declare no competing interests.

\bibliographystyle{elsarticle-num}
\bibliography{main}

\begin{thebibliography}{10}
\expandafter\ifx\csname url\endcsname\relax
  \def\url#1{\texttt{#1}}\fi
\expandafter\ifx\csname urlprefix\endcsname\relax\def\urlprefix{URL }\fi
\expandafter\ifx\csname href\endcsname\relax
  \def\href#1#2{#2} \def\path#1{#1}\fi

\bibitem{castells2011rise}
M.~Castells, The rise of the network society, Vol.~12, John Wiley \& Sons,
  2011.

\bibitem{unitedurbanization2018}
U.~Nations, 2018 revision of world urbanization prospects (2018).

\bibitem{travers1969experimental}
J.~Travers, S.~Milgram, An experimental study of the small world problem,
  Sociometry ((1969)) 425--443.

\bibitem{ugander2011anatomy}
J.~Ugander, B.~Karrer, L.~Backstrom, C.~Marlow, The anatomy of the facebook
  social graph, arXiv preprint arXiv:1111.4503 ((2011)).

\bibitem{backstrom2012four}
L.~Backstrom, P.~Boldi, M.~Rosa, J.~Ugander, S.~Vigna, Four degrees of
  separation, in: Proceedings of the 4th Annual ACM Web Science Conference,
  June 22 - 24, 2012, (2012), pp. 33--42.

\bibitem{edunov2016three}
S.~Edunov, C.~Diuk, I.~O. Filiz, S.~Bhagat, M.~Burke, Three and a half degrees
  of separation, Research at Facebook ((2016)).

\bibitem{bettencourt2013origins}
L.~M. Bettencourt, The origins of scaling in cities, Science 340~(6139)
  ((2013)) 1438--1441.

\bibitem{bettencourt2007growth}
L.~M. Bettencourt, J.~Lobo, D.~Helbing, C.~K{\"u}hnert, G.~B. West, Growth,
  innovation, scaling, and the pace of life in cities, Proceedings of the
  national academy of sciences 104~(17) ((2007)) 7301--7306.

\bibitem{carlino2007urban}
G.~A. Carlino, S.~Chatterjee, R.~M. Hunt, Urban density and the rate of
  invention, Journal of Urban Economics 61~(3) ((2007)) 389--419.

\bibitem{pickard2011time}
G.~Pickard, W.~Pan, I.~Rahwan, M.~Cebrian, R.~Crane, A.~Madan, A.~Pentland,
  Time-critical social mobilization, Science 334~(6055) ((2011)) 509--512.

\bibitem{khatib2011crystal}
F.~Khatib, F.~DiMaio, S.~Cooper, M.~Kazmierczyk, M.~Gilski, S.~Krzywda,
  H.~Zabranska, I.~Pichova, J.~Thompson, Z.~Popovi{\'c}, et~al., Crystal
  structure of a monomeric retroviral protease solved by protein folding game
  players, Nature structural \& molecular biology 18~(10) ((2011)) 1175--1177.

\bibitem{lazer2009life}
D.~Lazer, D.~Brewer, N.~Christakis, J.~Fowler, G.~King, Life in the network:
  the coming age of computational social, Science 323~(5915) (2009) 721--723.

\bibitem{eddington2019snowden}
P.~Eddington,
  \href{https://www.justsecurity.org/64464/the-snowden-effect-six-years-on/}{{The
  Snowden Effet, Six Years On}} (2019).
\newline\urlprefix\url{https://www.justsecurity.org/64464/the-snowden-effect-six-years-on/}

\bibitem{lapowsky2019cambridge}
I.~Lapowsky,
  \href{https://www.wired.com/story/cambridge-analytica-facebook-privacy-awakening/}{{How
  Cambridge Analytica Sparked the Great Privacy Awakening}} (2019).
\newline\urlprefix\url{https://www.wired.com/story/cambridge-analytica-facebook-privacy-awakening/}

\bibitem{EDPS2019}
E.~D.~P. Supervisor,
  \href{https://edps.europa.eu/sites/edp/files/publication/19-12-19_edps_proportionality_guidelines2_en.pdf}{Edps
  guidelines on assessing the proportionality of measures that limit the
  fundamental rights to privacy and the protection of personal data} (December
  2019).
\newline\urlprefix\url{https://edps.europa.eu/sites/edp/files/publication/19-12-19_edps_proportionality_guidelines2_en.pdf}

\bibitem{venkatadri1investigating}
G.~Venkatadri, E.~Lucherini, P.~Sapiezynski, A.~Mislove, Investigating sources
  of pii used in facebook’s targeted advertising, Proceedings on Privacy
  Enhancing Technologies 1 (2018) 18.

\bibitem{barabasi1999emergence}
A.-L. Barab{\'a}si, R.~Albert, Emergence of scaling in random networks, Science
  286~(5439) ((1999)) 509--512.

\bibitem{watts1998collective}
D.~J. Watts, S.~H. Strogatz, Collective dynamics of 'small-world' networks,
  nature 393~(6684) ((1998)) 440--442.

\bibitem{erdos1960evolution}
P.~Erdos, A.~R{\'e}nyi, On the evolution of random graphs, Publ. Math. Inst.
  Hung. Acad. Sci 5~(1) (1960) 17--60.

\bibitem{cadwalladr2018revealed}
C.~Cadwalladr, E.~Graham-Harrison, Revealed: 50 million facebook profiles
  harvested for cambridge analytica in major data breach, The Guardian 17
  (2018).

\bibitem{facebookprivacy2018}
M.~Schroepfer,
  \href{https://newsroom.fb.com/news/2018/04/restricting-data-access/}{An
  update on our plans to restrict data access on facebook} (April 2018).
\newline\urlprefix\url{https://newsroom.fb.com/news/2018/04/restricting-data-access/}

\bibitem{facebookfine2019}
\href{https://www.theguardian.com/technology/2019/jul/24/facebook-to-pay-5bn-fine-as-regulator-files-cambridge-analytica-complaint}{Facebook
  to pay \$5bn fine as regulator settles cambridge analytica complaint} (July
  2019).
\newline\urlprefix\url{https://www.theguardian.com/technology/2019/jul/24/facebook-to-pay-5bn-fine-as-regulator-files-cambridge-analytica-complaint}

\bibitem{barabasi2016network}
A.-L. Barab{\'a}si, et~al., Network Science, Cambridge university press, 2016.

\bibitem{BCG2014}
J.~Rose, C.~Barton, R.~Souza, J.~Platt, Data privacy by the numbers ((2014)).

\bibitem{mccarthy2002usa}
M.~T. McCarthy, {USA PATRIOT} act (2002).

\bibitem{franceinter2018}
\href{https://www.franceinter.fr/emissions/dans-le-pretoire/dans-le-pretoire-27-avril-2018}{Au
  coeur de l'antenj, la cellule des écoutes téléphoniques judiciaires}
  (April).
\newline\urlprefix\url{https://www.franceinter.fr/emissions/dans-le-pretoire/dans-le-pretoire-27-avril-2018}

\bibitem{homeland_security}
J.~Kelly, Enforcement of the immigration laws to serve the national interest,
  {Department of Homeland Security} (February (2017)).

\bibitem{mayer2016evaluating}
J.~Mayer, P.~Mutchler, J.~C. Mitchell, Evaluating the privacy properties of
  telephone metadata, Proceedings of the National Academy of Sciences 113~(20)
  ((2016)) 5536--5541.

\bibitem{aiello2000}
W.~Aiello, F.~Chung, L.~Lu, A random graph model for massive graphs, in: STOC,
  Vol. 2000, Citeseer, 2000, pp. 1--10.

\bibitem{flashlight2017}
H.~Domanski, This flashlight app for android wants to steal all your money,
  \url{http://www.techradar.com/news/this-flashlight-app-for-android-wants-to-steal-all-your-money},
  accessed: 2017-10-31 ((2017)).

\bibitem{stopczynski2014measuring}
A.~Stopczynski, V.~Sekara, P.~Sapiezynski, A.~Cuttone, M.~M. Madsen, J.~E.
  Larsen, S.~Lehmann, Measuring large-scale social networks with high
  resolution, PloS one 9~(4) ((2014)) e95978.

\bibitem{humbert2019survey}
M.~Humbert, B.~Trubert, K.~Huguenin, A survey on interdependent privacy, ACM
  Computing Surveys (CSUR) 52~(6) (2019) 1--40.

\bibitem{floridi2014open}
L.~Floridi, Open data, data protection, and group privacy, Philosophy \&
  Technology 27~(1) ((2014)) 1.

\bibitem{Mislove2010}
A.~Mislove, B.~Viswanath, K.~P. Gummadi, P.~Druschel,
  \href{http://doi.acm.org/10.1145/1718487.1718519}{You are who you know:
  Inferring user profiles in online social networks}, in: Proceedings of the
  Third ACM International Conference on Web Search and Data Mining, WSDM '10,
  February 04 - 06, 2010, (2010), pp. 251--260.
\newblock \href {https://doi.org/10.1145/1718487.1718519}
  {\path{doi:10.1145/1718487.1718519}}.
\newline\urlprefix\url{http://doi.acm.org/10.1145/1718487.1718519}

\bibitem{lampinen2015networked}
A.~Lampinen, Networked privacy beyond the individual: four perspectives
  to'sharing', in: Proceedings of The Fifth Decennial Aarhus Conference on
  Critical Alternatives, Aarhus University Press, (2015), pp. 25--28.

\bibitem{li2007t}
N.~Li, T.~Li, S.~Venkatasubramanian, t-closeness: Privacy beyond k-anonymity
  and l-diversity, in: Data Engineering, 2007. ICDE 2007. IEEE 23rd
  International Conference on, IEEE, (2007), pp. 106--115.

\bibitem{narayanan2005obfuscated}
A.~Narayanan, V.~Shmatikov, Obfuscated databases and group privacy, in:
  Proceedings of the 12th ACM conference on Computer and communications
  security, November 07 - 11, 2005, (2005), pp. 102--111.

\bibitem{callaway2000network}
D.~S. Callaway, M.~E. Newman, S.~H. Strogatz, D.~J. Watts, Network robustness
  and fragility: Percolation on random graphs, Physical review letters 85~(25)
  (2000) 5468.

\bibitem{shang2011hop}
Y.~Shang, W.~Luo, S.~Xu, L-hop percolation on networks with arbitrary degree
  distributions and its applications, Physical Review E 84~(3) (2011) 031113.

\bibitem{malware2017android}
C.~McGoogan, Millions of android devices infected with malware in popular judy
  game,
  http://www.telegraph.co.uk/technology/2017/05/31/millions-android-devices-infected-malware-popular-judy-game/,
  accessed: 2017-10-31 ((2017)).

\bibitem{ftc_flashlight}
{Federal Trade Commission},
  \href{https://www.ftc.gov/news-events/press-releases/2014/04/ftc-approves-final-order-settling-charges-against-flashlight-app}{{FTC
  Approves Final Order Settling Charges Against Flashlight App Creator}}
  ((2014)).
\newline\urlprefix\url{https://www.ftc.gov/news-events/press-releases/2014/04/ftc-approves-final-order-settling-charges-against-flashlight-app}

\bibitem{tamoco2018}
R.~Manthorpe,
  \href{https://www.wired.co.uk/article/tamoco-sam-amrami-proximity-tracking-mwc}{Sam
  amrani tracks you in pret. and at starbucks. and down the pub} (February
  2018).
\newline\urlprefix\url{https://www.wired.co.uk/article/tamoco-sam-amrami-proximity-tracking-mwc}

\bibitem{tau2020nextstep}
B.~Tau,
  \href{https://www.wsj.com/articles/next-step-in-government-data-tracking-is-the-internet-of-things-11606478401}{Next
  step in government data tracking is the internet of things}, The Wall Street
  Journal (November 2020).
\newline\urlprefix\url{https://www.wsj.com/articles/next-step-in-government-data-tracking-is-the-internet-of-things-11606478401}

\bibitem{bergman2022pegasus}
R.~Bergman, M.~Mazzetti,
  \href{https://www.nytimes.com/2022/01/28/magazine/nso-group-israel-spyware.html}{The
  battle for the world’s most powerful cyberweapon} (2022).
\newline\urlprefix\url{https://www.nytimes.com/2022/01/28/magazine/nso-group-israel-spyware.html}

\bibitem{EuropeContactTracing2020}
{European Commission},
  \href{https://ec.europa.eu/commission/presscorner/detail/en/ip_20_670}{{Coronavirus:
  An EU approach for efficient contact tracing apps to support gradual lifting
  of confinement measures}} (April 2020).
\newline\urlprefix\url{https://ec.europa.eu/commission/presscorner/detail/en/ip_20_670}

\bibitem{jain2019unveil}
S.~Jain, E.~Bensaid, Y.-A. de~Montjoye, {UNVEIL: Capture and Visualise WiFi
  Data Leakages}, in: The World Wide Web Conference, ACM, 2019, pp. 3550--3554.

\bibitem{sapiezynski2015opportunities}
P.~Sapiezynski, R.~Gatej, A.~Mislove, S.~Lehmann, Opportunities and challenges
  in crowdsourced wardriving, in: Proceedings of the 2015 ACM Conference on
  Internet Measurement Conference, ACM, October 28 - 30, 2015, (2015), pp.
  267--273.

\bibitem{sapiezynski2015tracking}
P.~Sapiezynski, A.~Stopczynski, R.~Gatej, S.~Lehmann, Tracking human mobility
  using wifi signals, PloS one 10~(7) ((2015)) e0130824.

\bibitem{android2022}
\href{{https://source.android.com/docs/core/connect/wifi-mac-randomization-behavior\#non-persistent}}{{MAC
  Randomization Behavior: Non-persistent randomization}} (2022).
\newline\urlprefix\url{{https://source.android.com/docs/core/connect/wifi-mac-randomization-behavior\#non-persistent}}

\bibitem{troncoso2020decentralized}
C.~Troncoso, M.~Payer, J.-P. Hubaux, M.~Salath{\'e}, J.~Larus, E.~Bugnion,
  W.~Lueks, T.~Stadler, A.~Pyrgelis, D.~Antonioli, et~al., Decentralized
  privacy-preserving proximity tracing, arXiv preprint arXiv:2005.12273 (2020).

\bibitem{dehaye2020proximity}
P.-O. Dehaye, J.~Reardon, Proximity tracing in an ecosystem of surveillance
  capitalism, in: Proceedings of the 19th Workshop on Privacy in the Electronic
  Society, 2020, pp. 191--203.

\bibitem{sapiezynski2019interaction}
P.~Sapiezynski, A.~Stopczynski, D.~D. Lassen, S.~Lehmann, Interaction data from
  the copenhagen networks study, Scientific Data 6~(1) (2019) 1--10.

\bibitem{yang2012network}
Y.~Yang, J.~Wang, A.~E. Motter, Network observability transitions, Physical
  Review Letters 109~(25) (2012) 258701.

\bibitem{hasegawa2013observability}
T.~Hasegawa, T.~Takaguchi, N.~Masuda, Observability transitions in correlated
  networks, Physical Review E 88~(4) (2013) 042809.

\bibitem{facebook2014}
A.~Smith,
  \href{http://www.pewresearch.org/fact-tank/2014/02/03/what-people-like-dislike-about-facebook/}{What
  people like and dislike about facebook} (February 2014).
\newline\urlprefix\url{http://www.pewresearch.org/fact-tank/2014/02/03/what-people-like-dislike-about-facebook/}

\bibitem{facebookusers2014}
Statista,
  \href{https://www.statista.com/statistics/247614/number-of-monthly-active-facebook-users-worldwide/}{Number
  of monthly active facebook users in the united states and canada as of 4th
  quarter 2018 (in millions).} (2018).
\newline\urlprefix\url{https://www.statista.com/statistics/247614/number-of-monthly-active-facebook-users-worldwide/}

\bibitem{newman2003structure}
M.~E. Newman, The structure and function of complex networks, SIAM review
  45~(2) (2003) 167--256.

\end{thebibliography}

\section*{Supplemental Experimental Procedures}

\counterwithin{figure}{section}
\renewcommand\thefigure{S\arabic{figure}}
\setcounter{figure}{0}  

\subsection*{\textbf{S1}: Alternatives to uniform node intrusions}

In this paper, we study edge- and node-observability through uniform node-based intrusions. While this assumption is realistic for the three use cases we present, other non-uniform attacks would be interesting to investigate.
Most of our mathematical definitions do not rely on this uniformity assumption and the framework can be easily extended with different compromised nodes distribution or targeting. For instance, targeted attacks to observe the entire network using as few nodes as possible have been studied before~\cite{yang2012network, hasegawa2013observability}. These differ from our work in that the attacker knows the structure of entire network, and chooses nodes to compromise in order to learn all of the nodes' private states (e.g. in a power grid).

\subsection*{\textbf{S2}: Facebook's node-observability}

Our analysis of Facebook's node-observability is based on the degree distribution of Facebook users in the United States in 2011~\cite{ugander2011anatomy}. We shift and truncate this distribution so that the average degree matches the 2014's average degree (which has been published \cite{facebook2014}). Formally, let $P_{2011}(d)$ be the distribution of degrees in 2011, such that $\sum_{d=0}^{5000} P_{2011}(d) = 1$. This distribution is truncated at 5000 because Facebook doesn't allow a user to have more than that number of friends. We compute an estimate degree distribution for 2014 as:
$$\widehat{P}_{2014}(d) = \left\{
    \begin{array}{ll}
        0 & \text{ if } d < \sigma \\
        A \cdot P_{2011}(d - \sigma) & \text{ otherwise}
    \end{array}
\right.$$
here the shift $\sigma$ is chosen such that the average degree $\sum_{d=0}^{5000} d \widehat{P}_{2014}(d)$ matches the empirical value, and the scaling constant $A$ is such that $\widehat{P}_{2014}$ sums to $1$. This is a simple approximation, assuming that users uniformly increased their friends count between 2011 and 2014.

We convert the distribution $\widehat{P}_{2014}$ to absolute degree counts by multiplying by $N=205 \cdot 10^6$, the number of Facebook users in the USA \cite{facebookusers2014}.
We then use theorem 3 to compute the 1-hop node-observability from the degree counts.
Our analysis shows that compromising $n_C = 270,000$ profiles, as Cambridge Analytica did, allows an attacker to observe 68.0 million users, giving a node-observability of 0.318.

\begin{figure}[H]
    \centering
    \includegraphics[width=10cm]{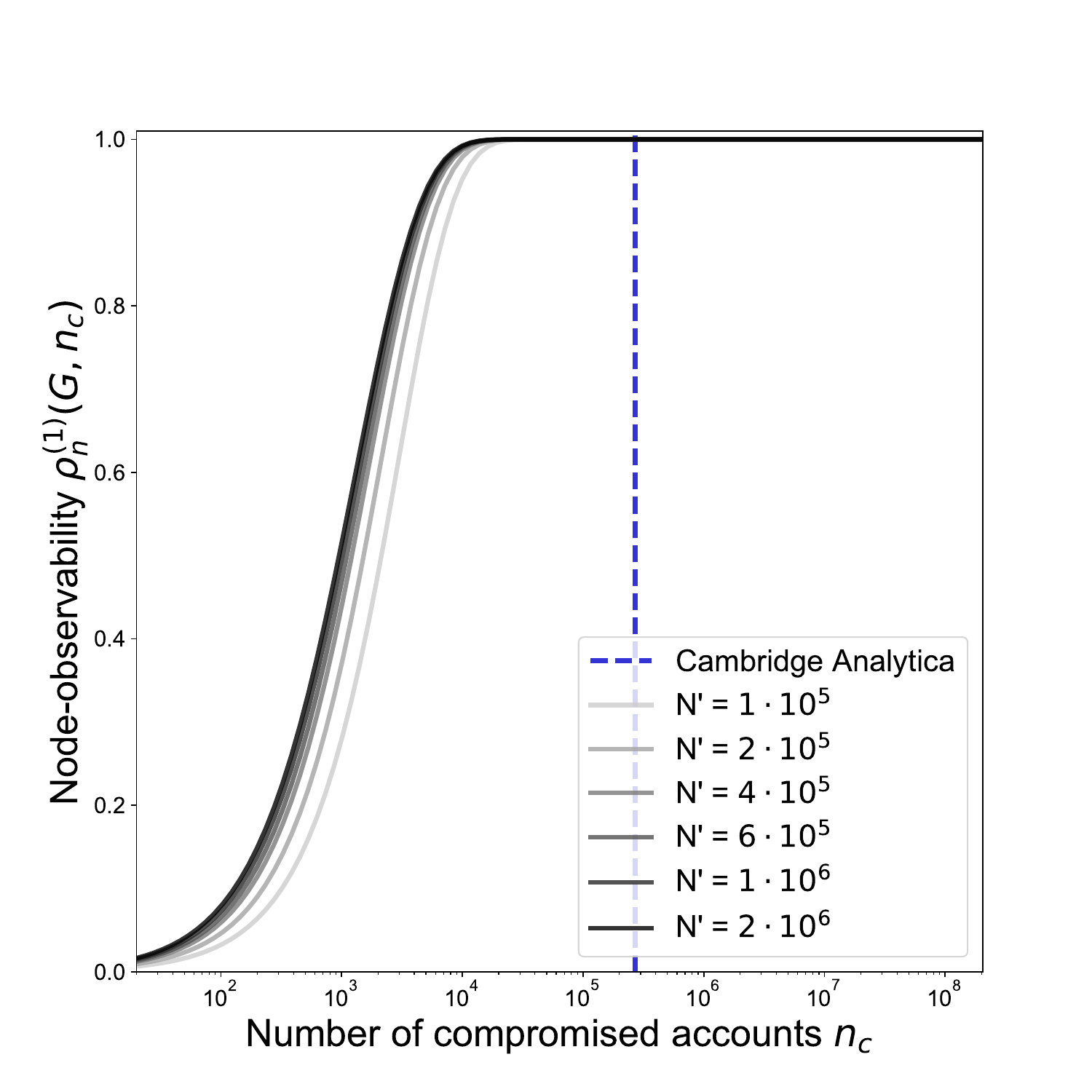}
    \caption{\textbf{Estimation of Cambridge Analytica's attack success if Facebook allowed for 2-hop}. 2-hop node-observability of graphs generated using the configuration model with degree distribution similar to that of the shifted Facebook degree distribution $\hat{P}_{2014}$ (solid curve). The curves correspond to graphs with different number of nodes $N'$ smaller than $N = 205 \cdot 10^6$, the number of Facebook users in the USA in 2014. The curves quickly converge for growing $n'$, and all agree for $n_p = 270,000$ (Cambridge Analytica).}
    \label{fig:appendix_2hopgno_fb}
\end{figure}

To study the 2-hops observability of the Facebook graph, we generate synthetic graphs from the distribution $\hat{P}_{2014}$ using the configuration model~\cite{newman2003structure}.
Generating such a graph with $N = 205 \cdot 10^6$ nodes is impractical, we compute the observability curve for graphs with $N' < N$ nodes, with $N'$ ranging from $10^5$ to $2 \cdot 10^6$, shown in Fig. \ref{fig:appendix_2hopgno_fb}.
Our results show that compromising a fraction $\frac{n_{CA}}{\text{US population}} = \frac{270 000}{205 \cdot 10^6} \approx 0.0013$ of the network would leads to a node-observability of $1.0$ on all the synthetic graphs. This analysis strongly suggests that a 2-hop policy would have enabled Cambridge Analytica to observe the whole Facebook network.

\subsection*{\textbf{S3}: Close proximity Networks}

\newcommand{\mygraph}{\mathbb{G}}

The first step of our computation of the observability of cities is to estimate the average probability of observation from a fixed number of sensors in a $1~km^2$ cell. We use Algorithm~\ref{algo:sampling} to estimate this from geo-tagged hourly co-location data, which we model as a function of the hourly time-stamp (in a set of hours $\mathcal{T}$) and the cell (from a set of $1~km^2$ cells $\mathcal{C}$) to the set of all graphs (denoted by $\mathbb{G}$) $\mygraph:\mathcal{T}\times\mathcal{C} \rightarrow \mathbb{G}$.

\begin{algorithm}[H]
\caption{Average Probability of Observation from hourly co-location data}
\begin{algorithmic}[1]
\Procedure{AvgProbaObs}{$n_p, \mygraph, n_{samples}$}
\State $observed \gets 0$
\For {$i = 1, \dots, n_{samples}$}
    \State Pick a cell uniformly at random $c \gets \mathcal{C}$;
    \State Pick a time uniformly at random $h \gets \mathcal{T}$;
    \State Define $G(c, h) = (V, E)$, pick $n_p$ nodes uniformly at random from $V$, $V_p$;
    \State Pick a target $u$ uniformly at random from non-infected nodes $u \gets V \backslash V_p$;
    \If {there is an edge between $u$ and a node in $V_p$}
        \State $observed += 1$
    \EndIf
    \State $m_i \gets x \cdot B_i$
\EndFor
\State \textbf{return} $observed/n_{samples}$
\EndProcedure
\end{algorithmic}
\label{algo:sampling}
\end{algorithm}

\begin{figure}[H]
    \centering
    \includegraphics[width=.5\textwidth]{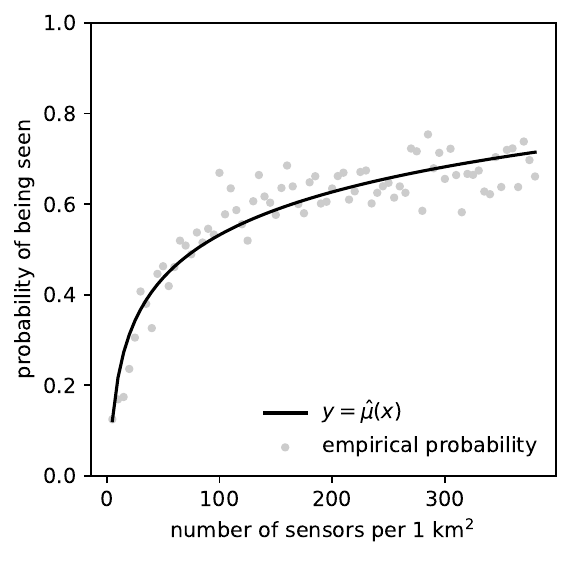}
    \caption{\textbf{Average probability of observation in hourly co-location graphs in a $1~km^2$ cell.} Each dot is obtained from \textsf{AvgProbaObs} on the DTU dataset, using $n_{samples} = 25$. The solid curve is $\avgprobafit(n_p) = 0.13 \log(n_p) - 0.05$, which fits the observed data~($R^2 = 0.876$).}
    \label{fig:suppmat_DTU_fit}
\end{figure}

We use procedure \textsf{AvgProbaObs} with $n_p$ ranging from 5 to 385, in increments of 5, with $n_{samples} = 25$ (about 10000 total samples). Fig.~\ref{fig:suppmat_DTU_fit} presents our results, along with the fit by $\avgprobafit(n_p) = 0.13 \log(n_p) - 0.05$~($R^2 = 0.876$).

\end{document}